\documentclass[conference]{IEEEtran}
\IEEEoverridecommandlockouts
\usepackage{amsmath,amssymb,amsfonts}
\usepackage{algorithmic}
\usepackage{textcomp}
\usepackage{xcolor}
\usepackage{graphicx}
\usepackage{amssymb}
\usepackage{booktabs}
\usepackage{listings}
\usepackage{csquotes}
\usepackage{subcaption}
\usepackage{comment}

\usepackage[ruled,vlined,linesnumbered]{algorithm2e}

\SetCommentSty{mycommfont}

\SetAlFnt{\small}
\SetAlCapFnt{\small}
\SetAlCapNameFnt{\small}
\SetAlCapHSkip{0pt}
\IncMargin{-\parindent}

\lstdefinestyle{CPP}{
frame=single,  breaklines=true, basicstyle=\scriptsize,
numbers=left, numberstyle=\tiny, stepnumber=1, numbersep=5pt,%
backgroundcolor=\color{gray!10},%
}%

\begin{document}

\title{Customizing Pareto Simulated Annealing for Multi-objective Optimization of Control Cabinet Layout\thanks{This research has received funding from the Swedish Knowledge Foundation under Grant No. 20150088}
}

\author{
	\IEEEauthorblockN{Sabri Pllana}
	\IEEEauthorblockA{
		\textit{Linnaeus University}\\
		V\"{a}xj\"{o}, Sweden \\
		sabri.pllana@lnu.se}
	\and
    \IEEEauthorblockN{Suejb Memeti}
	\IEEEauthorblockA{
		\textit{Link\"{o}ping University}\\
		Link\"{o}ping, Sweden \\
		suejb.memeti@liu.se}
	\and
	\IEEEauthorblockN{Joanna Kolodziej}
	\IEEEauthorblockA{
		\textit{Cracow University of Technology}\\
		Cracow, Poland \\
		jokoldziej@pk.edu.pl}
}


\IEEEspecialpapernotice{\footnotesize(Preprint, CSCS22, \copyright 2019 IEEE)}

\maketitle

\begin{abstract}
Determining  the  optimal  location  of  control cabinet  components  requires the exploration of a large configuration space. For real-world control cabinets it is impractical to evaluate all possible cabinet configurations. Therefore, we need to apply methods for intelligent exploration of cabinet configuration space that enable to find a near-optimal configuration without evaluation of all possible configurations.
In this paper, we describe an approach for multi-objective optimization of control cabinet layout that is based on Pareto Simulated Annealing. Optimization aims at minimizing the total wire length used for interconnection of components and the heat convection within the cabinet. We simulate heat convection to study the warm air flow within the control cabinet and determine the optimal position of components that generate heat during the operation. We evaluate and demonstrate the effectiveness of our approach empirically for various control cabinet sizes and usage scenarios.  
\end{abstract}


\section{Introduction}
\label{sec:introduction}

\begin{figure}
	\centering
		\includegraphics[width=0.4\textwidth]{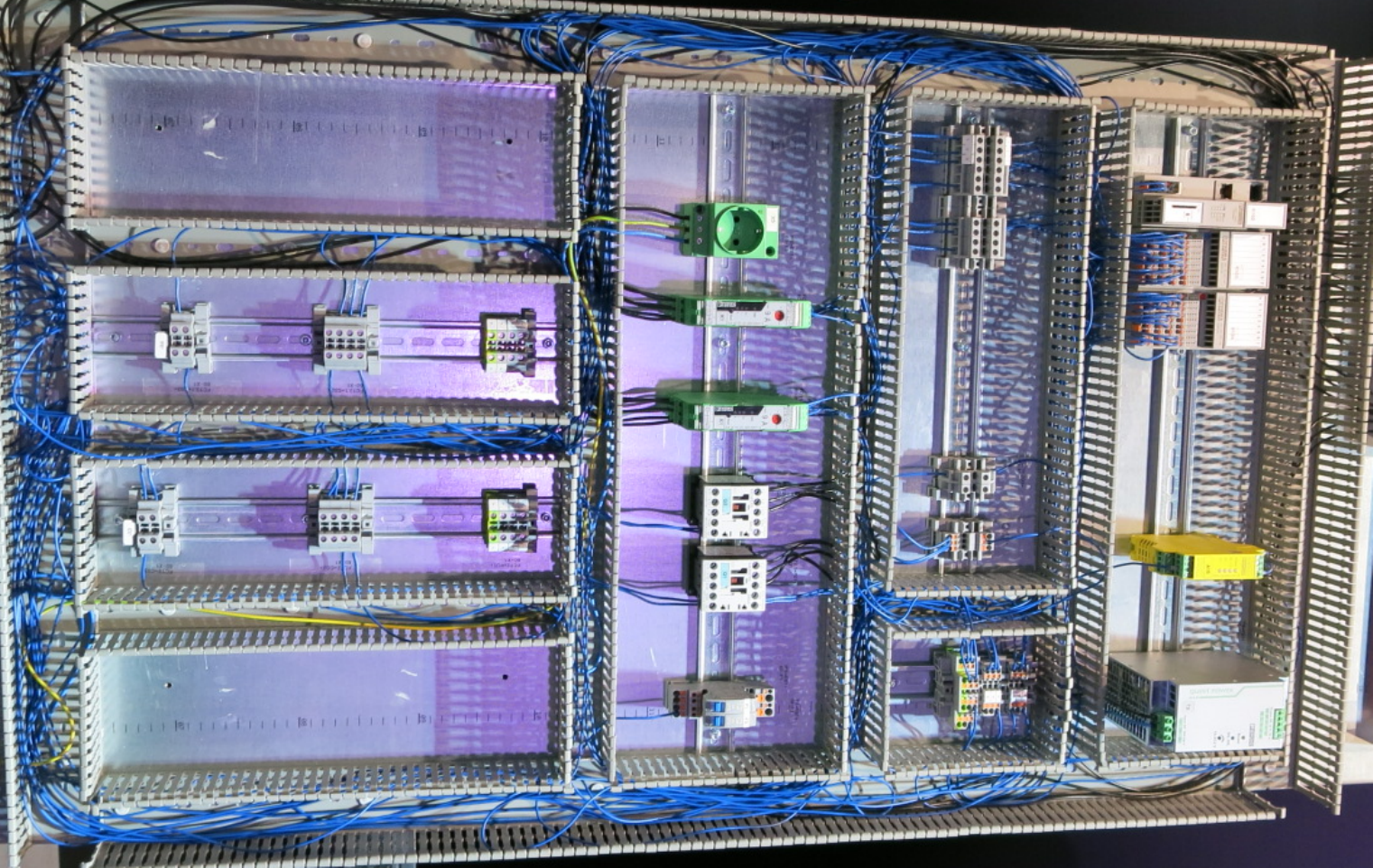}
	\caption{An example of control cabinet with components mounted on standardized metal rails known as DIN rails.}
    \label{fig:cabinet-example}
\end{figure}

A control cabinet manages a system (such as, an industrial robotic arm) and contains a collection of interconnected electronic and electrical components. Figure \ref{fig:cabinet-example} depicts an example of control cabinet that is under construction. We may observe that components are mounted on standardized metal rails that are also known as DIN rails. Many companies today assemble control cabinets manually, from reading and interpreting the electrical circuit of a control cabinet, through \emph{planning of the control cabinet layout}, to actual assembly of components and cables. This process could be significantly improved if some of these steps (such as, the cabinet layout planning) are automated. For instance, layout planning of a control cabinet of moderate size usually takes about two hours of the engineer's time. 

The functionality of a control cabinet is described as electrical circuit using abstract symbols to represent electrical components and lines to represent the wires that interconnect components. An electrical circuit is an abstract representation of control cabinet that does not contain the information about the physical properties, such as, component dimensions. To produce the control cabinet, one needs first to develop a real-world cabinet layout based on electrical circuit. In the cabinet layout (1) abstract symbols are  mapped to real-world components with their corresponding physical properties, and (2) location of components within the cabinet is determined. 

The search space of all possible control cabinet configurations is very large, and therefore a brute force approach that would evaluate all possible solutions is impractical. Moreover, the search space is a \emph{discrete} cabinet configuration space and the \emph{gradient} or \emph{downhill} methods are not applicable \cite{Press:2007}. The Simulated Annealing \cite{Kirkpatrick:1983} is a combinatorial optimization method that deserves our attention, also because of the capability of finding the global optimum including cases that have many local optimums. Considering the fact that many single-objective combinatorial optimization problems are NP-hard, it is reasonable to expect that the corresponding multi-objective versions are usually harder to solve \cite{Lukata:1994,Grzonka:2018,Khan:2010}. \emph{Pareto Simulated Annealing (PSA)} is an extension of the Simulated Annealing for efficient multi-objective combinatorial optimization of complex systems \cite{Czyzak:1998,Teghem:2000,Duh:2007}.

The Simulated Annealing is widely used for combinatorial optimization of complex systems, such as, decentralized scheduling in Grid computing environments \cite{Pop:2008}, optimization of DNA sequence analysis on heterogeneous computing systems \cite{Memeti:2017}, gate assignment problem in the context of an airport \cite{Drexl:2008}, furniture arrangement \cite{Yu:2011}, or hybrid vehicle routing \cite{Yu:2017}. 

In this paper, we describe our approach for customization of the multi-objective Simulated Annealing method for optimization of control cabinet layout. Objectives of the optimization procedure are \emph{heat convection} and the total \emph{wire length} used for interconnection of control cabinet components. We simulate heat convection to study the warm air flow within the control cabinet and determine the optimal position of components that generate heat during the operation. Moreover, we describe the experimental results with respect to optimization of objectives and the program execution time for various cabinet sizes. For a cabinet with 14 components, trying about 4.7\% of all feasible configurations requires about 6.4 seconds and results with average 2x improvement of the wire-length in addition to the average 1.6x improvement of the heat-level. We study the interactive use of our implementation of control cabinet layout optimization for adapting the layout when a component is replaced. The assumption is that the component that is replaced has the same functionality, but it may have different properties (that is, width, height, depth, heat) that may lead to invalidation of the previously optimized cabinet layout and the cabinet layout re-optimization is required. While an experienced engineer typically needs hours to plan the layout of a moderate size control cabinet, our approach is able to generate a near-optimal layout within few seconds.

Major contributions of this paper include: 

\begin{itemize}
	\item Simulation of heat convection within a control cabinet to illustrate the flow of warm air and determine the optimal position of components that generate heat during the operation. 
	\item Customization of multi-objective Simulated Annealing method for combinatorial optimization of control cabinet layout.
	\item Experimental evaluation of our approach using control cabinets with 14, 21, 41 components. 
\end{itemize}

The rest of this paper is structured as follows. Section \ref{sec:problem} introduces the problem of control cabinet layout optimization that is addressed in this paper. We describe our approach for multi-objective optimization of control cabinet layout in Section \ref{sec:approach}. An empirical evaluation of the proposed method is described in Section \ref{sec:evaluation}. We discuss the related work in Section \ref{sec:rw}. Section \ref{sec:summary} concludes this paper.

\section{Problem Statement}
\label{sec:problem}

We describe a control cabinet as a collection of interconnected components. Properties of cabinet components include: component number, ID, width, height, depth, list of component numbers to which the component is connected to, and the component heat indicator. An example of control cabinet description is provided in Table \ref{tab:initial-description}. 

In this paper, the optimization process of the control cabinet layout aims at minimizing the total wire length used for interconnection of control cabinet components while placing as far as it is possible at the top components that during the operation generate significant amounts of heat.

\begin{figure}
	\centering
    \begin{subfigure}{0.22\textwidth}
		\includegraphics[width=\textwidth]{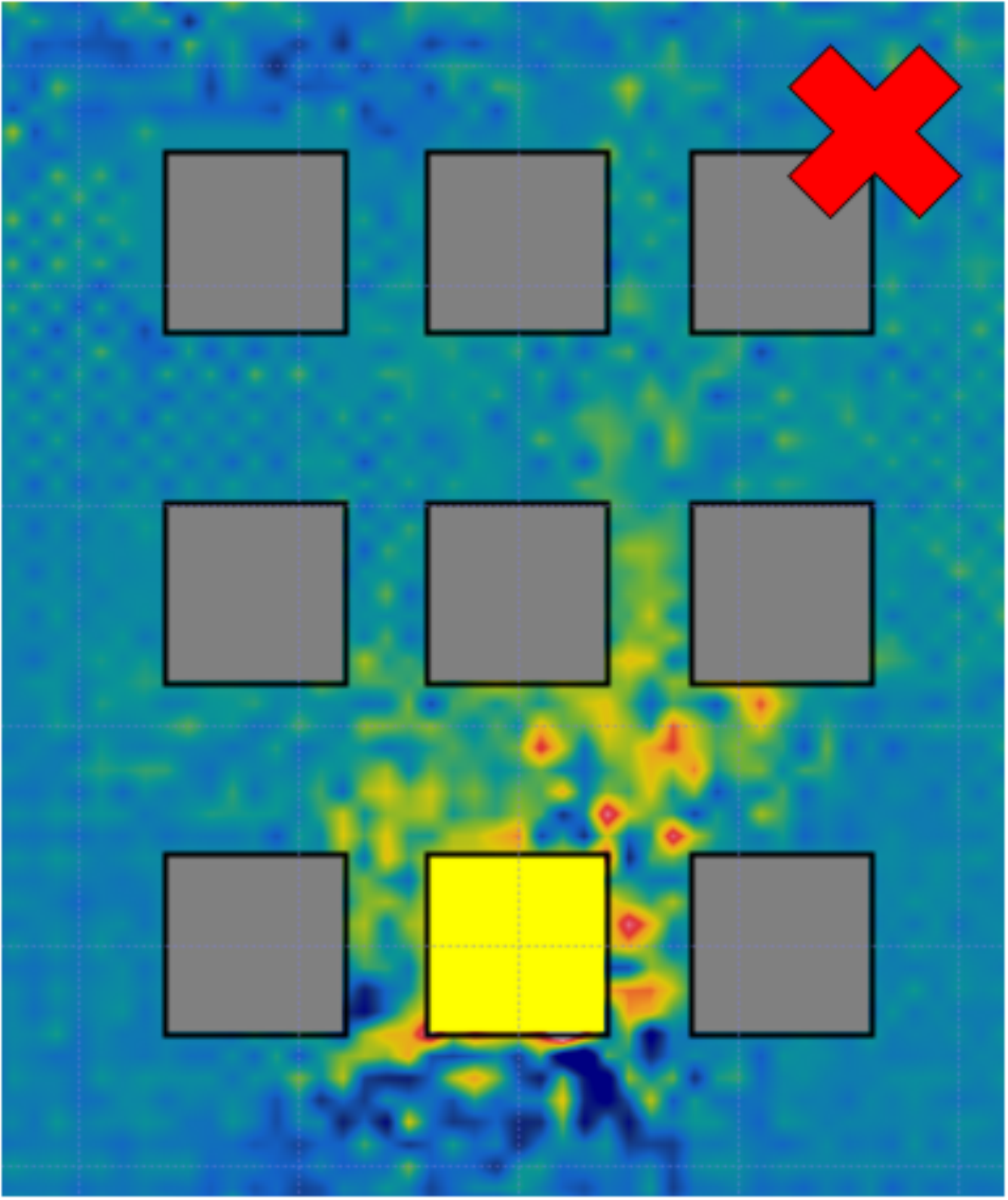}
		\caption{hot component at bottom}
		\label{fig:heat-sim-bad}
	\end{subfigure}
	\hfill
	\begin{subfigure}{0.22\textwidth}
		\includegraphics[width=\textwidth]{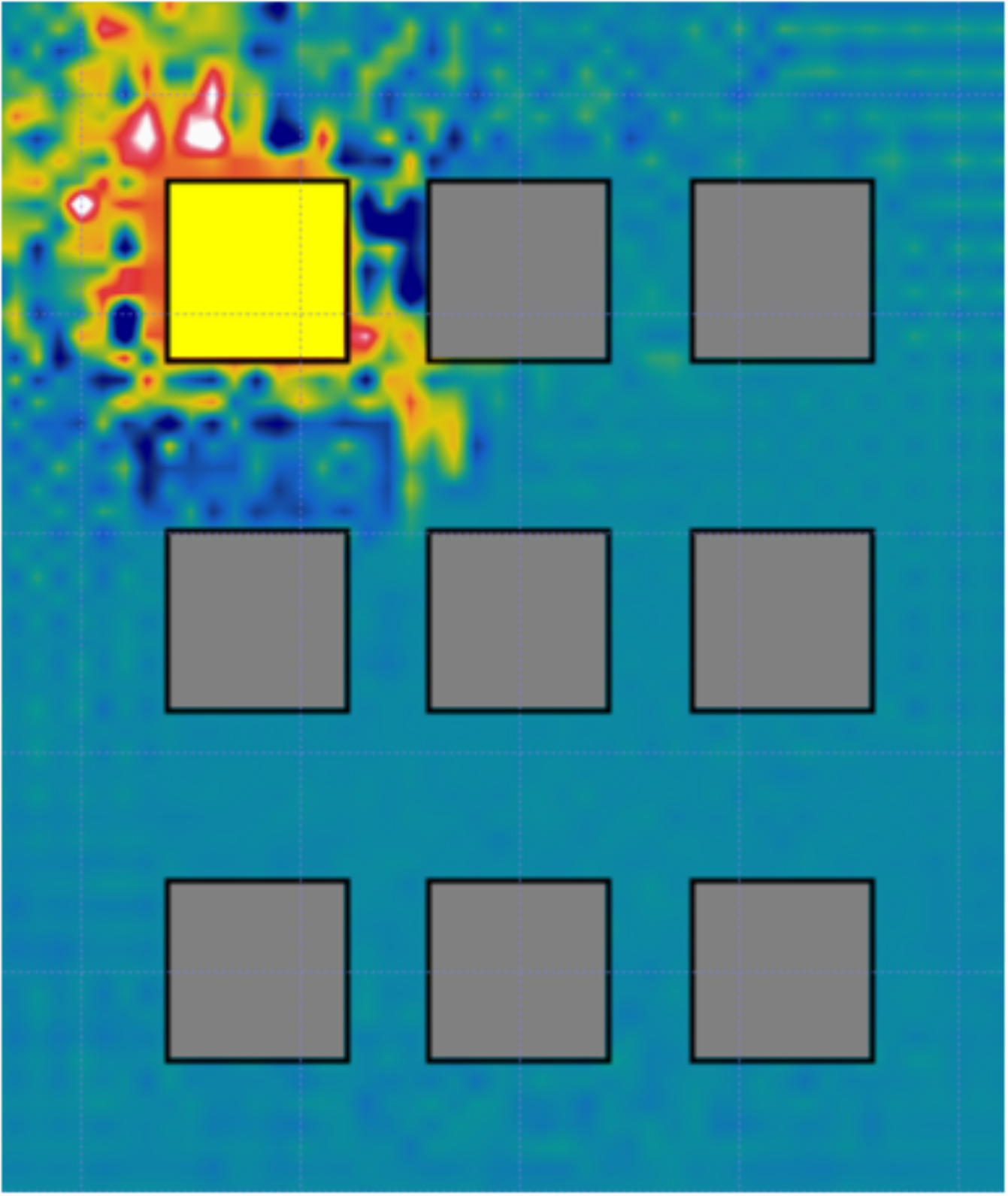}
		\caption{hot component at the top}
		\label{fig:heat-sim-good}
	\end{subfigure}	
	\caption{Heat convection simulation of a hypothetical cabinet layout; to avoid heating other components it is not recommended to place at the bottom of the cabinet components that during the operation generate significant amounts of heat.}
    \label{fig:heat-sim}
\end{figure}

Figure \ref{fig:heat-sim} depicts an example of two-dimensional heat convection  simulation with Energy2D \cite{Energy2D} of a hypothetical cabinet layout. Placing a component that during the operation dissipates heat at the bottom of the cabinet may significantly contribute to the increase of the temperature of the whole cabinet and will heat unnecessary also the components that are placed above it (Figure \ref{fig:heat-sim-bad}). It is recommended that hot components are placed at the top of the cabinet (Figure \ref{fig:heat-sim-good}) to optimize the flow of the warm air within the cabinet.

\section{Customization of Simulated Annealing for Optimization of Control Cabinet Layout}
\label{sec:approach}

In this section, we first introduce basic concepts of the Simulated Annealing. Thereafter, we describe how we customize the multi-objective Simulated Annealing in the context of our approach for control cabinet layout optimization.

Combinatorial analysis is a mathematical discipline that studies arrangements of a finite number of discrete objects \cite{Nemhauser:1988,Lawler:2001,Press:2007}. While initially the scientists were concerned with enumeration or existence of arrangements of objects, the focus is now in finding the optimal arrangement of objects (that is in our case cabinet components). In the context of combinatorial optimization, the minimum or maximum is sought of an \emph{objective function} (also known as \emph{cost function}) that depends on many independent variables. The objective function reflects the configuration of components of the system under study, and the outcome of the function represents the \emph{goodness} of the system. A minimum or maximum of a cost function is known as \emph{extremum}. A \emph{local extremum} is the minimum or maximum of the cost function within a limited range of function variables, while the \emph{global extremum} applies to the whole range of values of function variables. 

The Simulated Annealing method has demonstrated the capability of finding the global optimum including cases of cost functions that have many local optimums. Simulated Annealing was introduced by Kirkpatrick et al. \cite{Kirkpatrick:1983} for solving combinatorial optimization problems (such as, the Traveling Salesman Problem). The method is inspired by statistical mechanics, which is a physics discipline that studies features of a large number of atoms. The thermal equilibrium of a configuration of atoms with locations $l_i$ at temperature $T$ is expressed using the Boltzmann probability distribution,

\begin{equation} \label{eq:boltzmann}
    P(E(l_i)) \propto exp(-E(l_i) / k_B T)
\end{equation}

where $k_B$ is the Boltzmann's constant that represents the relationship between temperature and energy. When the temperature of material in liquid (or melted) state is decreased slowly, the material may reach through a series of re-configurations its crystal state that is the \emph{minimum energy state} of the material; this process is also known as \emph{annealing} (that is strengthening) of material. 

A procedure for \emph{simulation of annealing} is provided by Metropolis et al. \cite{Metropolis:1953}. Initially a group of atoms is at energy state $E$. Then a new configuration of atoms is generated that results with an energy change of $\delta E$. If $\delta E$ is less than or equal zero, then the new configuration of atoms is accepted as a system improvement towards reaching the minimum energy state. In case that $\delta E$ is larger than zero, the acceptance decision is made in probabilistic fashion by using the Boltzmann probability distribution (Equation \ref{eq:boltzmann}). Basically, a random number $u$ is generated from the uniform random distribution $U(0,1)$ and is compared with $P(\delta E)$. If $u$ is less than $P(\delta E)$, the new configuration of atoms is accepted. This feature of Simulated Annealing that with a certain probability accepts also worse solutions (that is higher energy state configurations) enables to avoid local minimums in search for the global minimum of energy state. While the temperature $T$ is higher, it is more likely to accept worse solutions and get out of a local minimum, in favor of searching for a global minimum. The lower the temperature, less probable that it accepts new configurations with a higher energy state. 

\begin{table}[t]
	\centering
	\caption{Instantiating Simulated Annealing concepts for cabinet layout optimization.}
		\begin{tabular}{p{1.3cm}p{3cm}p{3cm}}
			\toprule
			Domain          & Statistical mechanics        & Electrical and electronics engineering\\
			\hline
			System          & Collection of atoms  & Control cabinet\\
			Elements        & Atoms                & Cabinet components\\
			Objective       & Energy               & Total wire length, heat\\
			Parameter       & Temperature          & Cabinet layout reconfiguration parameter\\
		    Plan            & Annealing schedule   & Cabinet layout reconfiguration plan\\
			\bottomrule
		\end{tabular}
	\label{tab:concepts}
\end{table}

Considering the fact that many single-objective combinatorial optimization problems are NP-hard, it is reasonable to expect that the corresponding multi-objective versions are usually harder to solve \cite{Lukata:1994}. An efficient approach for solving multi-objective optimization problems is to extend the Simulated Annealing for handling optimization of more than one objective; this method is known as the multi-objective Simulated Annealing or \emph{Pareto Simulated Annealing} \cite{Czyzak:1998,Teghem:2000,Duh:2007}. While single-objective Simulated Annealing searches for a single-solution, Pareto Simulated Annealing maintains a set $S$ of interacting solutions (also known as generating solutions) during the search procedure that is conceptually similar to Genetic Algorithms \cite{Michalewicz:1996}.

First, a set of generating solutions is initialized. If a new solution $s^{n}$ is not dominated by the previous solution $s$, then $s^{n}$ is added to the set of potentially efficient solutions $M$. Solutions that are dominated by $s^{n}$ are removed from $M$. If the new solution $s^{n}$ is not better than the previous solution $s$, the probability to accept $s^{n}$ is as follows,

\begin{equation} \label{eq:psa}
    P(s,s^{n},T,\Lambda^s) \propto min \{1, exp(\sum_{\omega = 1}^{\Omega}(\lambda_\omega^s(f_\omega(s) - f_\omega(s^{n})) / T)\}
\end{equation}

In Equation \ref{eq:psa}, $\omega$ is one of the $\Omega$ objectives considered for optimization, $f_\omega$ is the objective function for objective $\omega$, $\Lambda^s$ is the vector of utilized weights $\lambda_\omega^s$ for solution $s$. Weights of objectives are varied during the iterative search to explore the whole set of non dominated solutions (that is, the set of efficient solutions or the Pareto front). A higher weight associated with objective $\omega$ implies a higher probability of improving $\omega$. The sum of all objective weights is equal to one. 

Table \ref{tab:concepts} maps basic concepts of the Simulated Annealing method from the original domain of statistical mechanics to the domain of electrical and electronics engineering that is relevant for this paper. A collection of atoms is mapped to control cabinet, atoms are mapped to cabinet components, objective energy is mapped to objectives total wire length and heat. Annealing temperature and schedule are adapted to the probabilistic procedure of selecting cabinet layout configurations. 

Algorithm \ref{alg:psa} depicts the pseudo-code of the Pareto Simulated Annealing that is tailored for control cabinet layout optimization with respect to the heat convection and total length of wire needed to connect the components within the cabinet. The algorithm start by initiating the temperature and cooling schedule, and randomly generating an initial set of cabinet configurations (see Line \ref{alg:psa-init} - \ref{alg:psa-conf-gen}). Then, for each of the generated cabinet configurations, we check for dominance (that is, check if the heat level or the wire-length is better), and update the set of best configurations $M$; all solutions removed from $M$ that have either overall greater heat-level or wire-length, and the new cabinet configuration $s^{n}$ is added. Lines \ref{alg:psa-start-loop} - \ref{alg:psa-cool-down} show the main loop of the algorithm, which is executed until the temperature has decreased to or below 1. Then we loop around all cabinet configurations in $S$, and for each element $s$, we generate a new solution $s^{n}$ (line \ref{alg:psa-gen-sol}), we check if it dominates $s$ and update the set of best cabinet configurations accordingly (line \ref{alg:psa-check-dominance-gen-sol}). Thereafter, we update the weights for each objective of $s$ (line \ref{alg:psa-update-weight-start} - \ref{alg:psa-update-weight-stop}). At the first iteration, we do set the weight for each objective randomly (line \ref{alg:psa-set-weight}). The probability function $P(s,s^{n},T,\Lambda^s)$ (Equation \ref{eq:psa}) is used to decide whether to update solution $s$ with the newly generated one $s^{n}$ (line \ref{alg:psa-boltzman}). Then we decrease the temperature according to the cooling schedule (line \ref{alg:psa-cool-down}).

\begin{algorithm}[t]
	\KwData{Cabinet components}
	\KwResult{Best cabinet configuration}
	set initial temperature $T_{SA}$, cooling schedule, and constant $c$ used to update weights\; \label{alg:psa-init}
	randomly generate an initial set of cabinet configurations S\; \label{alg:psa-conf-gen}
	
	\ForEach{$s \in S $} { \label{alg:psa-check-dominance}
	    check for dominance and update the set of best cabinet configurations $M$\;
	}
	
	\While{$T_{SA} > 1$}{ \label{alg:psa-start-loop}
    	\ForEach{$s \in S $} {
    		generate new cabinet configuration $s^{n}$\; \label{alg:psa-gen-sol}
    		
    		\uIf{$s^{n}$ dominates $s$}{\label{alg:psa-check-dominance-gen-sol}
    	    	update the set of best cabinet configurations $M$\; 	

        		\uIf{not the first iteration}{
        		    update objective weights such that:\; \label{alg:psa-update-weight-start}
        		    \uIf {objective of $s^{n} \geq$ objective of s} {
        		        multiply the corresponding objective weight by $c$\;
        		    }
        		    \Else {
        		        divide the corresponding objective weight by $c$\;
        		    }
        		    ... \tcp{Do the same for all objectives}
        		    update weights of s\; \label{alg:psa-update-weight-stop}
        		}
        		\Else{
        		    set random weight values\; \label{alg:psa-set-weight}
        		}
    		}
    		
    		use the probability function $P(s,s^{n},T,\Lambda^s)$ to decide whether to update $M$ with $s^{n}$ \tcp{see eq. \ref{eq:psa}} \label{alg:psa-boltzman}
    		
    		decrease $T_{SA}$\ according to the plan\; \label{alg:psa-cool-down}
    	}
	}
	\caption{A high-level description of the customized Pareto Simulated Annealing for cabinet placement optimization.}
	\label{alg:psa}
\end{algorithm}

\section{Evaluation}
\label{sec:evaluation}

In this section, we first describe the experimentation environment. Thereafter, we describe the results with respect to optimization and execution time. Furthermore, we illustrate the usefulness of our approach with a scenario of interactive use of our implementation for cabinet layout reconfiguration. 

\subsection{Experimentation Environment}
\label{sec:environment}

For evaluation of our approach, we consider three different scenarios (that is cabinets), and we name them A, B, and C. Table \ref{tab:cabinets} lists the number of (hot) components for each of the cabinets, the number of connections (that is, wires), the total wire length and heat level for their initial solution. Cabinet A is the simplest one, which has in total 14 components, four of which are hot, and in total 5 wires connecting the components with each other. The average values for heat level and total wire length of the generated initial solutions are 8.4 for the heat level, and 869.5 for the wire length. Cabinet B has 21 components, 6 of which are hot, and in total 22 wires connecting various components. The average heat level of the initial solutions for cabinet B is 18.1, whereas the average wire length is 4992.8. Cabinet C is the most complex one, which obviously results with the largest exploration space. In total, it has 40 components, 12 of which are hot and should be considered to be placed on top of the cabinet. There are 88 wires in total that connect various components. Please note that one component might be connected to one or more components. The average value for heat level of the initial solutions for cabinet C is 63.2, whereas the total wire length is 24817.6. Cabinet B is considered as the best representation of a real world scenario, however we use cabinet A and C to evaluate the scalability of our solution. 

\begin{table}[t]
	\caption{Characteristics of the considered cabinets for evaluation of our solution.}
	\label{tab:cabinets}
	\begin{tabular}{@{}p{3cm}p{1.5cm}p{1.5cm}p{1.5cm}@{}}
		\toprule
		& Cabinet A 	& Cabinet B 	& Cabinet C \\ \midrule
		\# of components     		& 14        		 & 21        		 & 41        \\
		\# of hot components 	  & 4         		   & 6         			& 12        \\
		\# of connections    		 & 5         		   & 22        		  & 88        \\
		initial heat level           			   & 8.4       			& 18.1      	   & 63.2      \\
		initial total wire length    			& 869.5     	   & 4992.8    	   & 24817.6   \\ \bottomrule
	\end{tabular}
\end{table}

In addition to the label that determines whether a component is hot or not, each of the components has details regarding their width and height, which are used to determine their correct placement in the cabinet and to measure the total wire length. 

We run our experiments in a notebook that comprises a 2.5 GHz Intel Core i7 CPU, 16 GB DDR3 RAM, running a MAC OS X 10.14 and Java Runtime Environment 1.8.0. We execute our experiments 10 times and report the average values. We use the initial temperature of the PSA algorithm to control the number of iterations (that is, the number of tried cabinet configurations). We vary the initial annealing temperature $T_{SA}$ between 100, 1000, and 10000. While increasing the temperature results with more optimal solution, the optimization process becomes more time consuming. 
\subsection{Results}
\label{sec:results}

\subsubsection{Improvement of Heat Level and Wire Length}

\begin{figure}[ht]
	\centering
    \begin{subfigure}{\linewidth}
		\includegraphics[width=\linewidth]{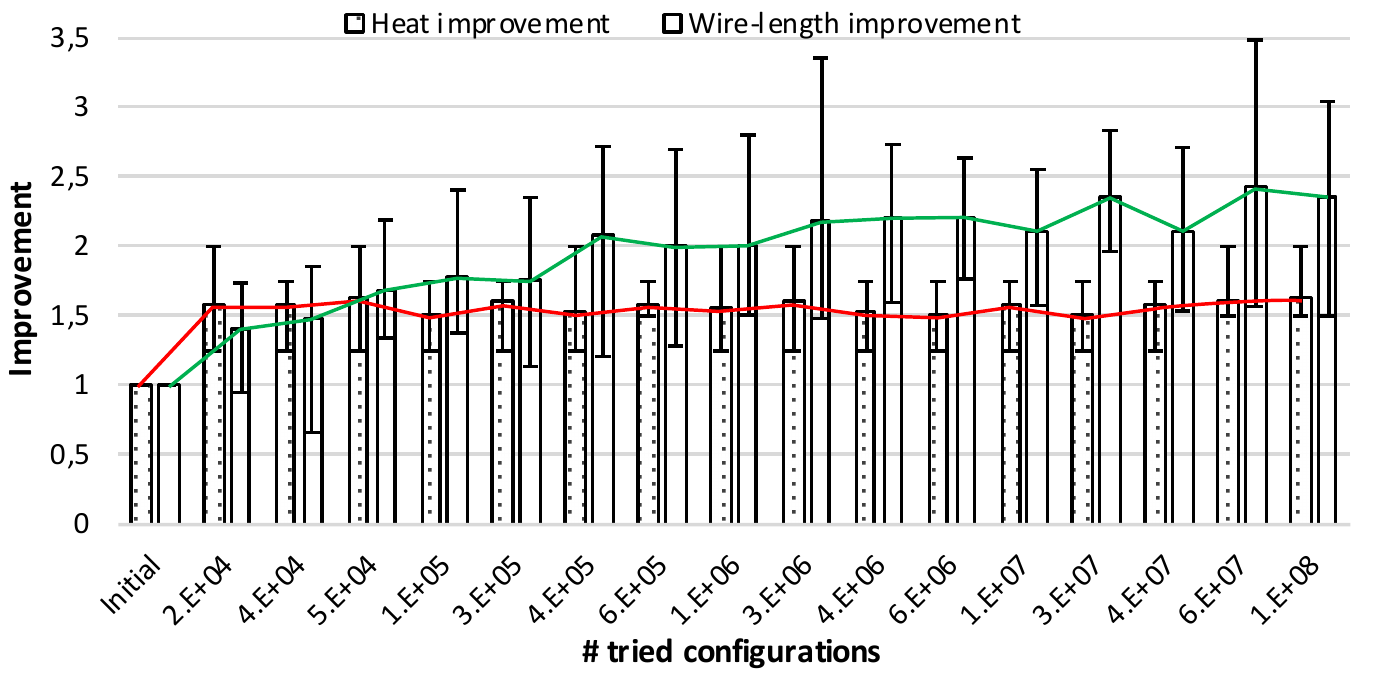}
		\caption{Cabinet A}
		\label{fig:cabinet-a-improv}
	\end{subfigure}
	\hfill
	\begin{subfigure}{\linewidth}
		\includegraphics[width=\linewidth]{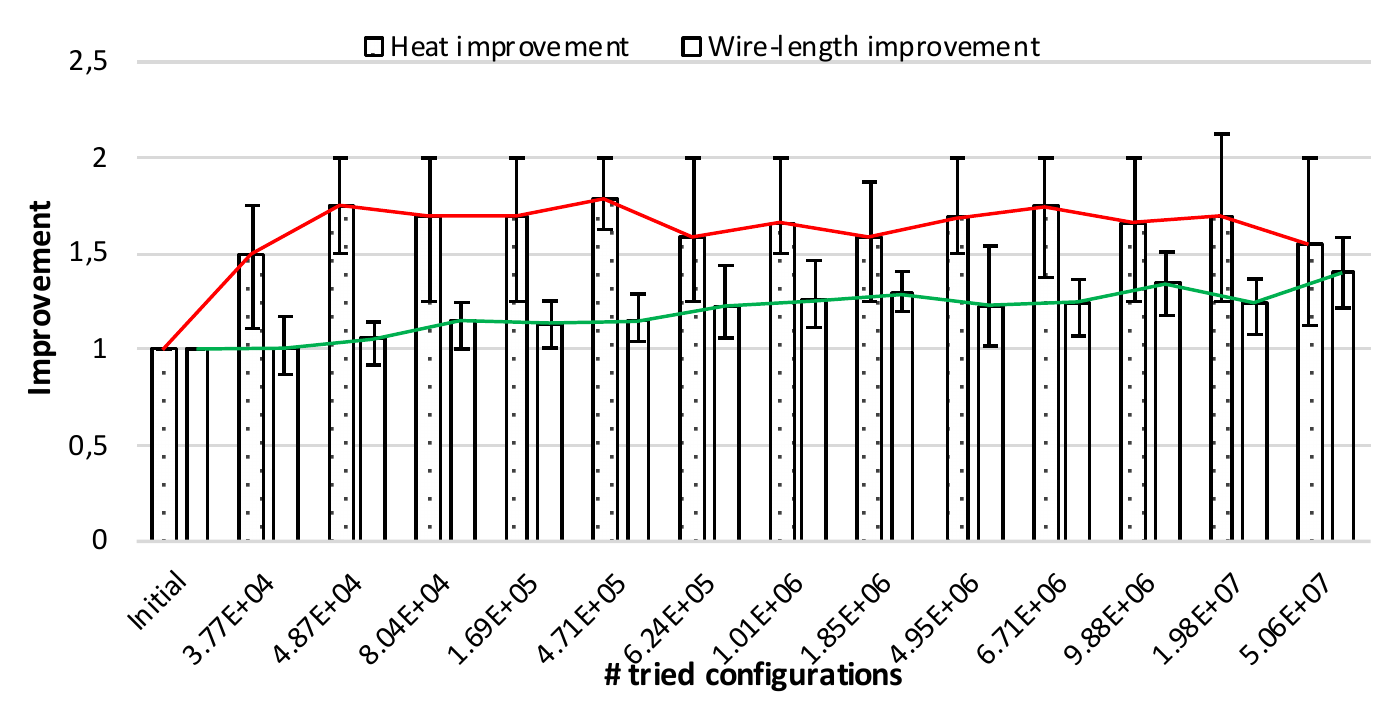}
		\caption{Cabinet B}
		\label{fig:cabinet-b-improv}
	\end{subfigure}
	\hfill
	\begin{subfigure}{\linewidth}
		\includegraphics[width=\linewidth]{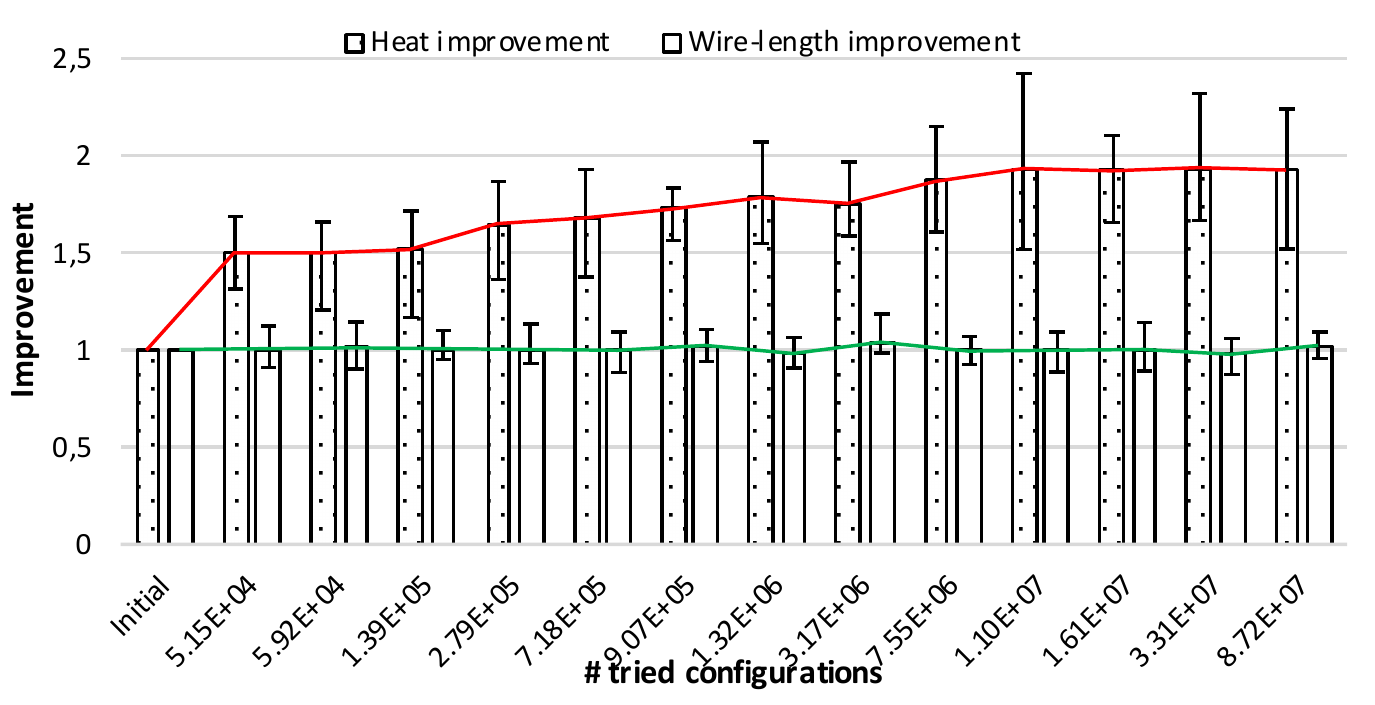}
		\caption{Cabinet C}
		\label{fig:cabinet-c-improv}
	\end{subfigure}
	\caption{Improvement of heat level and wire length compared to the initial solution for various cabinet configurations, and various number of iterations. We define the improvement as follows: Improvement($\omega$) = Initial($\omega$) / Final($\omega$), where $\omega$ is the objective (in our case heat level or wire length).}
    \label{fig:optimization-improvement}
\end{figure}

Figure \ref{fig:optimization-improvement} shows the improvement of the cabinet heat level and wire length compared to the initial solution. We use three various cabinet configurations (see Table \ref{tab:cabinets}), which comprise various numbers of components and component characteristics, such as type (hot or not hot), width, height, and connections with other components. We also vary the number of iterations by adjusting the temperature cooling rate of the PSA algorithm. Reducing the temperature cooling rate results with larger number of considered cabinet configurations, hence increasing the chances to find a more optimal solution. The positive and negative "error" bars on the charts indicate the observed maximal, respectively, minimal improvements observed out of the 10 runs for a particular number of iterations. For instance, for 6e+07 iterations used for optimization of cabinet A, the maximal observed improvement of the wire-length is 3.5x, whereas the minimal observed improvement is 1.5x. The average, indicated by the column bars, for this particular case is 2.4x. 

Since the algorithm is designed to prioritize the minimization of the cabinet heat first and then the wire length, we may observe that in all cases the algorithm first optimizes for cabinet heat, and when increasing the number of iterations the algorithm tries to optimize for wire length as well. For instance, in the first two cases (cabinet A in Fig. \ref{fig:cabinet-a-improv} and cabinet B in Fig. \ref{fig:cabinet-b-improv}) the optimal level of heat is reached with relatively low number of iterations (in case of cabinet A, after trying only 0.010\% of all possible configurations; in case of cabinet B, after trying only 8e-11\% of all possible configurations). Increase of the number of iterations, results with optimization of the wire length. 

In the case of the optimization of cabinet C in Fig. \ref{fig:cabinet-c-improv}, we may observe that by trying about 8.55e-37\% of all configurations is not enough for optimization of the cabinet for both heat and wire-length. While the heat level is certainly improving (that is reduced), there is no significant improvement of the wire-length required to connect the components with each other. Further increasing the number of iterations may result in improvement of the wire-length as well, however since the space is very large (about 1.02e+46 total configurations), increasing the number of cabinet configurations results with significant increase of the algorithm execution time. 

\subsubsection{Impact on the execution time}

\begin{figure}[ht]
	\centering
    \begin{subfigure}{\linewidth}
		\includegraphics[width=\linewidth]{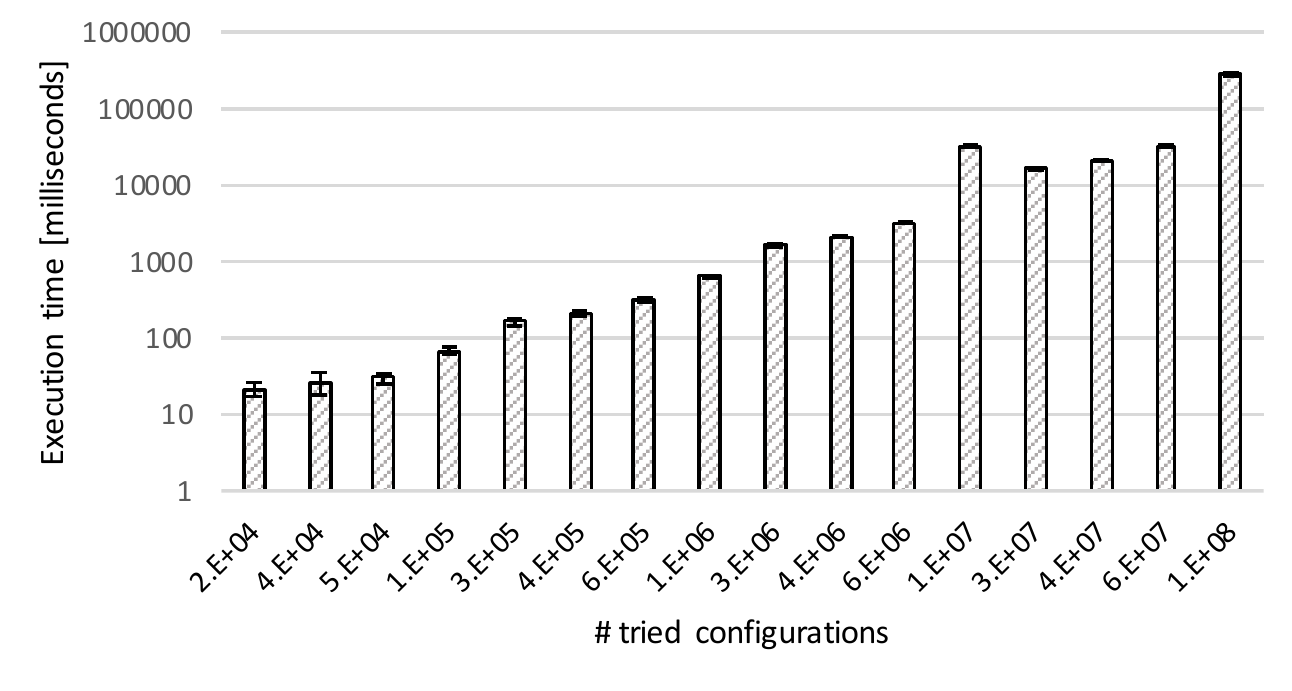}
		\caption{Cabinet A}
		\label{fig:cabinet-a-time}
	\end{subfigure}
	\hfill
	\begin{subfigure}{\linewidth}
		\includegraphics[width=\linewidth]{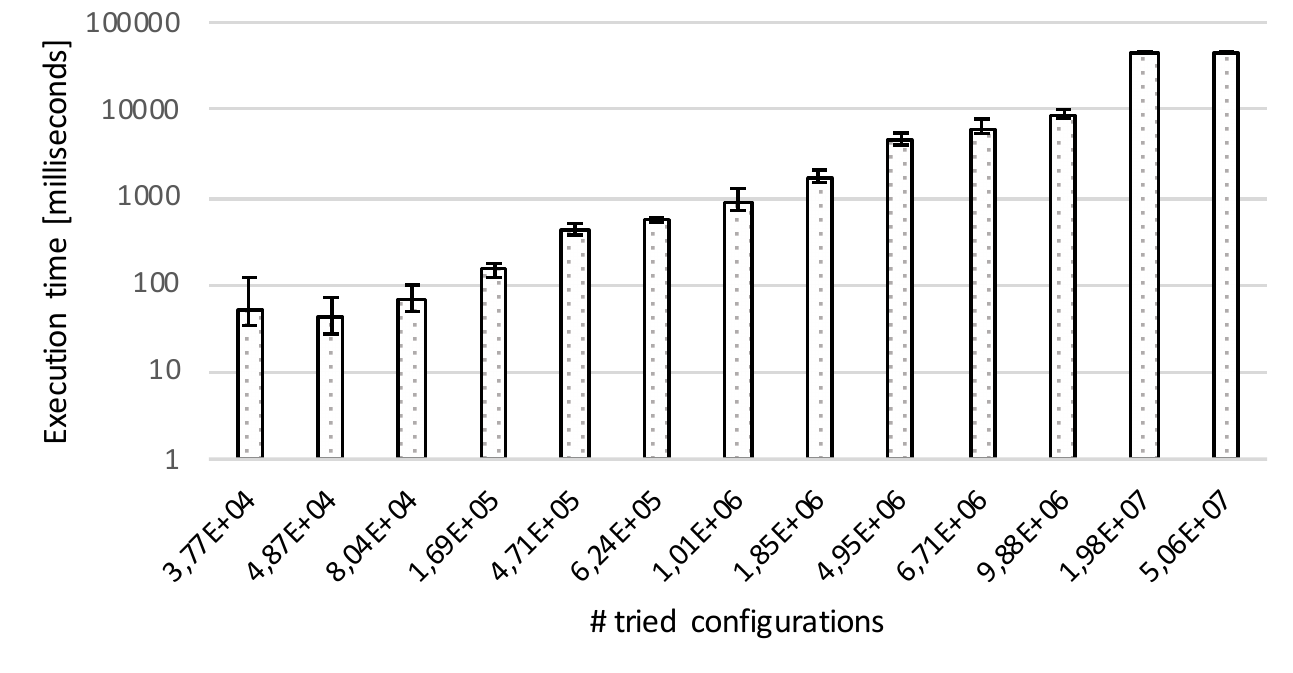}
		\caption{Cabinet B}
		\label{fig:cabinet-b-time}
	\end{subfigure}
	\hfill
	\begin{subfigure}{\linewidth}
		\includegraphics[width=\linewidth]{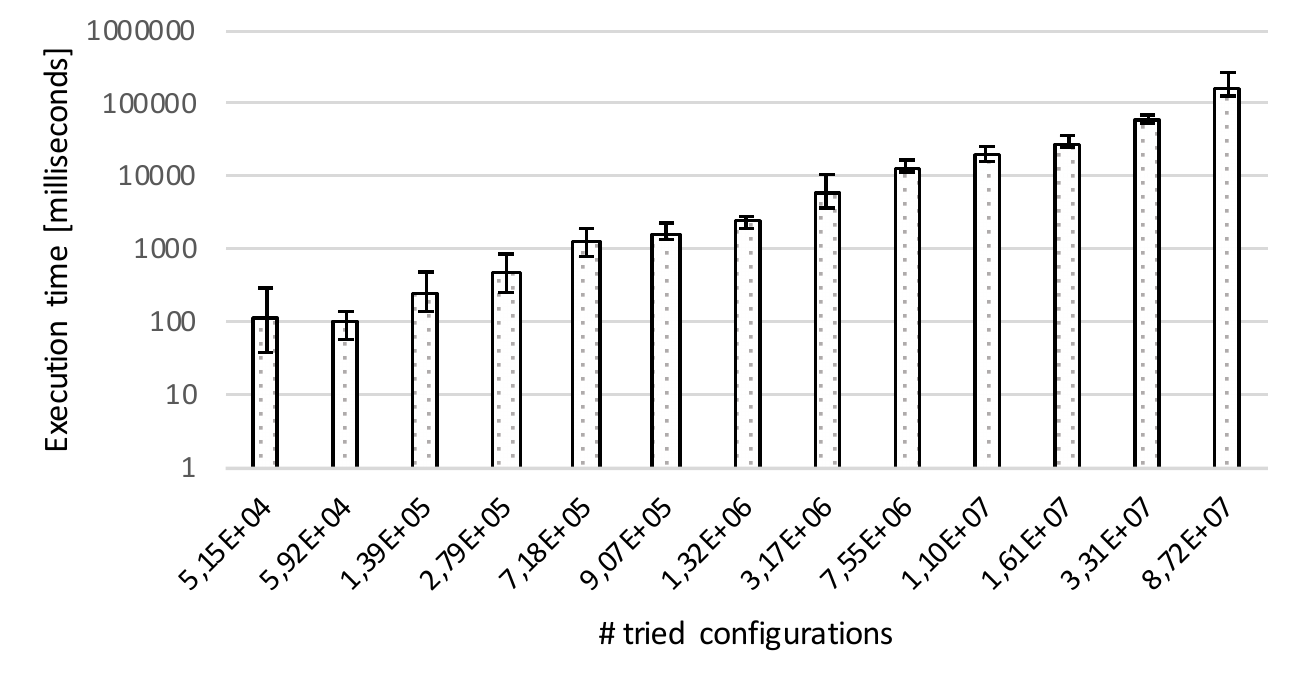}
		\caption{Cabinet C}
		\label{fig:cabinet-c-time}
	\end{subfigure}
	\caption{The execution time of the optimization process using the Pareto Simulated Annealing algorithm with various cabinet sizes and various number of iterations. As we increase the number of iterations (by adjusting the algorithm cooling ratio), the PSA algorithm tries more configurations (which results with more optimal solutions, see Fig. \ref{fig:optimization-improvement}),  and the total execution time increases.}
    \label{fig:optimization-execution-time}
\end{figure}

Figure \ref{fig:optimization-execution-time} depicts the execution time of the optimization process for the selected cabinet configurations (cabinet A, B, and C) and numbers of iterations. For all cases, we may observe that, as expected, there is a close relationship between the number of iterations and the execution time, which means that the increase of numbers of iterations results with increase of the execution time. For cabinet A (see Fig. \ref{fig:cabinet-a-time}, we were able to reach on average 1.57x and 1.4x improvement of the heat-level and wire-length, respectively, of the cabinet by trying only 0.01\% of configurations, which takes about 21 milliseconds only. Increasing the number of iterations (that is, tried configurations) to 0.12\% increases the execution time to 160 milliseconds and results with average 1.6x improvement for heat and 1.76x for wire length. Trying about 4.7\% of all configurations, requires about 6.4 seconds and results with average 2.12x improvement of the wire-length in addition to the average 1.6x improvement of the heat-level. 

For cabinet B however, due to the large space exploration, trying only 1.0e-07\% of all possible configurations takes about 45 seconds to finish. In this case, we were able to achieve an average of 1.55x improvement for heat and 1.4x for wire-length. 

Cabinets with larger number of components, such as cabinet C, has a much larger exploration space (about 1.02E+46 possible configurations). Only a limited number of configurations that can be tried is feasible. In our case, we run experiments by trying only up to 8.54e-37\% (that is 87'186'168) of all configurations. Execution of these experiments requires on average 158.7 seconds to finish. One way to improve the execution time is by utilizing high-performance parallel computing systems \cite{pllana17,peppher:2011,hpc:2015}, however this is out of scope of this paper.

\subsection{Cabinet reconfiguration}

In the section, we illustrate the interactive use of our implementation of control cabinet layout optimization for adapting the layout when a component is replaced. The assumption is that the component that is replaced has the same functionality, but it may have different properties (that is, width, height, depth, heat) that may lead to invalidation of the previously optimized cabinet layout and the cabinet layout reconfiguration is required.

Table \ref{tab:initial-description} lists components of a hypothetical control cabinet. This description is used as input to our implementation of multi-objective cabinet layout optimization. Properties of each cabinet component include: component number, ID, width, height, depth, list of component numbers to which the component is connected to, and the component heat indicator.

\begin{table}
\centering
\caption{Component description of a hypothetical control cabinet. This description is used as input to our implementation of multi-objective cabinet layout optimization.}
\begin{tabular}{rcccccc}
\toprule
    \# & ID   & Width & Height & Depth & Connects To & Is Hot \\
\hline
    1  & 0001 & 120.0 & 150.0  & 200.0 & {[}3{]}     & 1     \\
    2  & 0002 & 160.0 & 165.0  & 200.0 & {[}1{]}     & 1     \\
    3  & 0002 & 160.0 & 165.0  & 200.0 & {[}7{]}     & 0     \\
    4  & 0003 & 176.5 & 158.0  & 200.0 & {[}5{]}     & 0     \\
    5  & 0004 & 132.6 & 165.0  & 200.0 & {[}6{]}     & 1     \\
    6  & 0005 & 149.0 & 155.0  & 200.0 & {[}15{]}    & 0     \\
    7  & 0005 & 149.0 & 155.0  & 200.0 & {[}14{]}    & 0     \\
    8  & 0005 & 149.0 & 155.0  & 200.0 & {[}1;5{]}   & 0     \\
    9  & 0006 & 129.1 & 165.0  & 200.0 & {[}10{]}    & 0     \\
    10 & 0007 & 120.0 & 150.5  & 200.0 & {[}12{]}    & 0     \\
    11 & 0008 & 138.0 & 152.0  & 200.0 & {[}10{]}    & 0     \\
    12 & 0008 & 138.0 & 152.0  & 200.0 & {[}11{]}    & 0     \\
    13 & 0008 & 138.0 & 152.0  & 200.0 & {[}12{]}    & 0     \\
    14 & 0009 & 111.6 & 170.0  & 200.0 & {[}12;15{]} & 0     \\
    15 & 0010 & 121.3 & 150.0  & 200.0 & {[}11;6{]}  & 0     \\
\bottomrule
\end{tabular}
\label{tab:initial-description}
\end{table}

\begin{figure*}[ht]
	\centering
	\begin{subfigure}[b]{0.44\textwidth}
	\centering
	    \includegraphics[width=\linewidth]{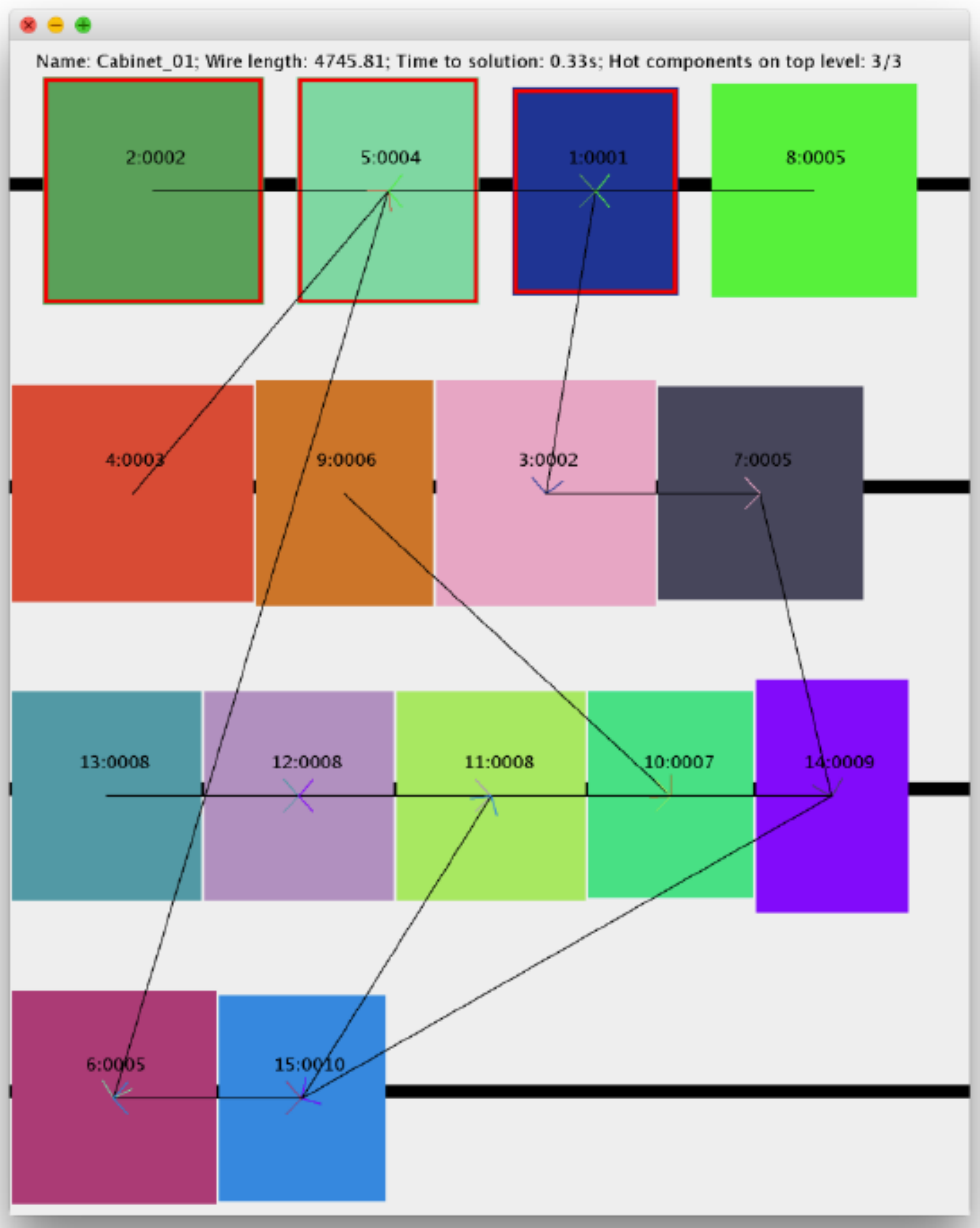}
    	\caption{Initial solution for the cabinet described in Table \ref{tab:initial-description}.}
    	\label{fig:initial-solution}
	\end{subfigure}
	\hfill
    \begin{subfigure}[b]{0.44\textwidth}
    \centering
		\includegraphics[width=\linewidth]{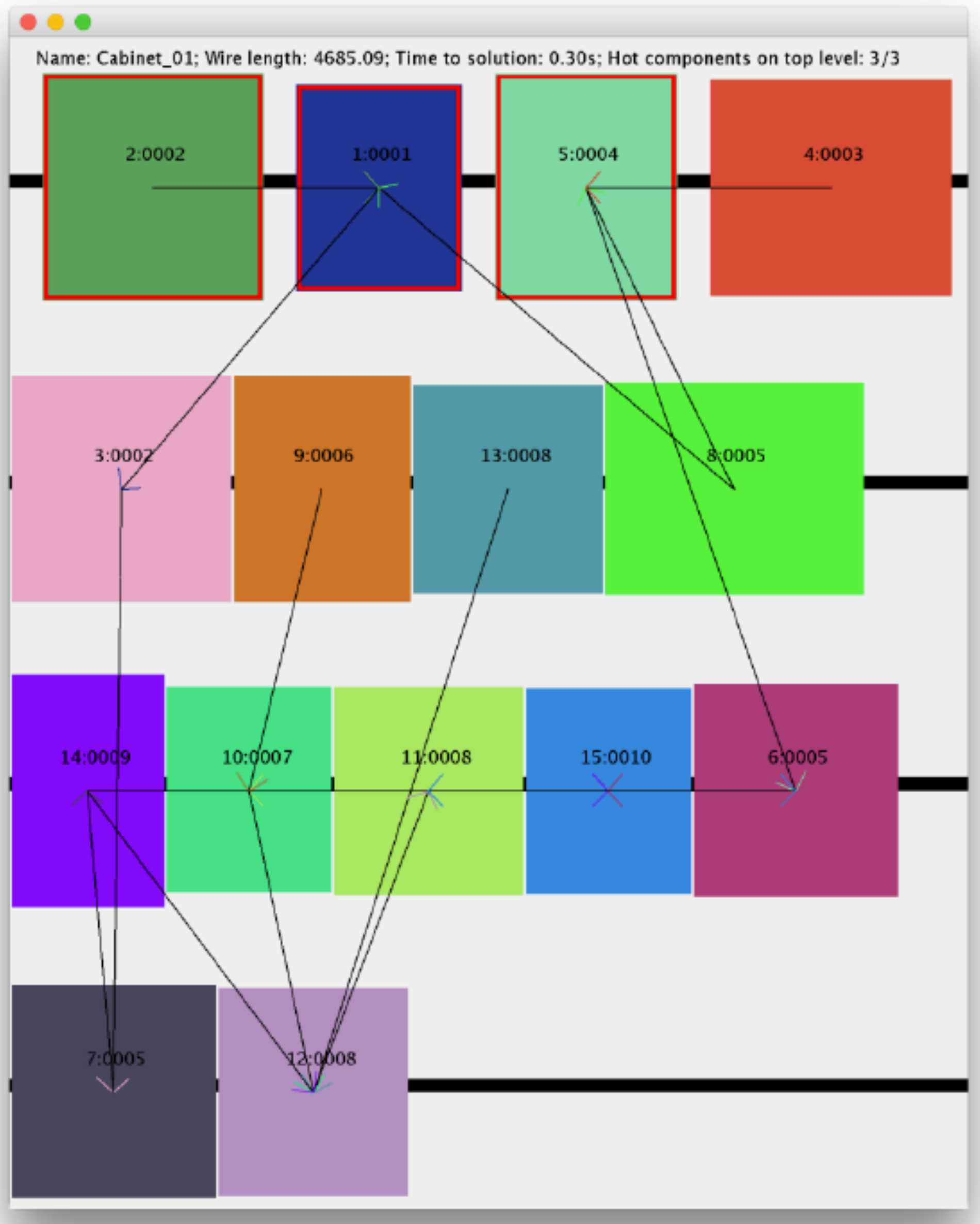}
		\caption{Component \#8 is replaced with a wider version.}
		\label{fig:wider-component}
	\end{subfigure}
	\hfill
    \begin{subfigure}[b]{0.44\textwidth}
    \centering
		\includegraphics[width=\linewidth]{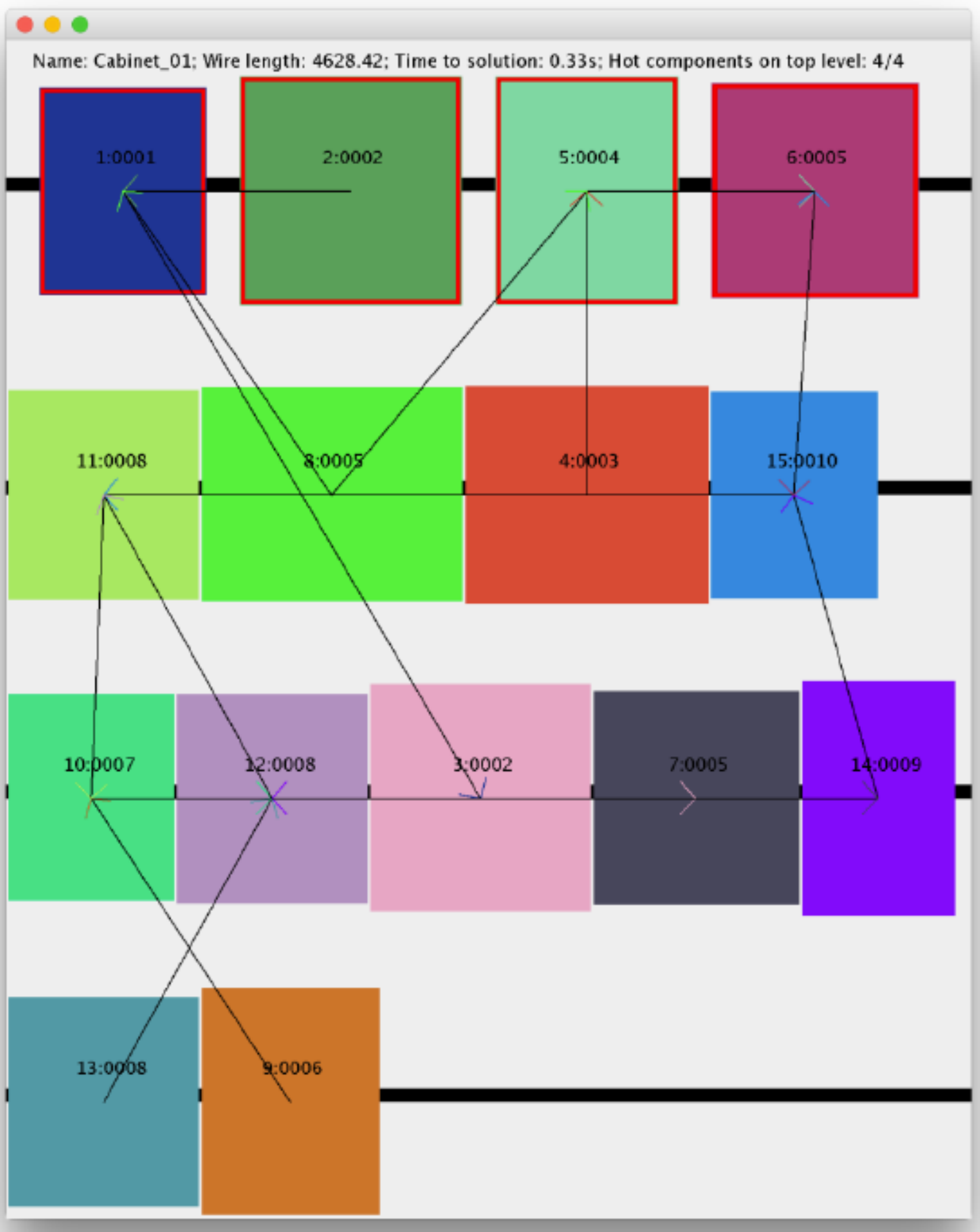}
		\caption{New component \#6 is hot.}
		\label{fig:hot-component}
	\end{subfigure}
	\hfill
	\begin{subfigure}[b]{0.44\textwidth}
	\centering
		\includegraphics[width=\linewidth]{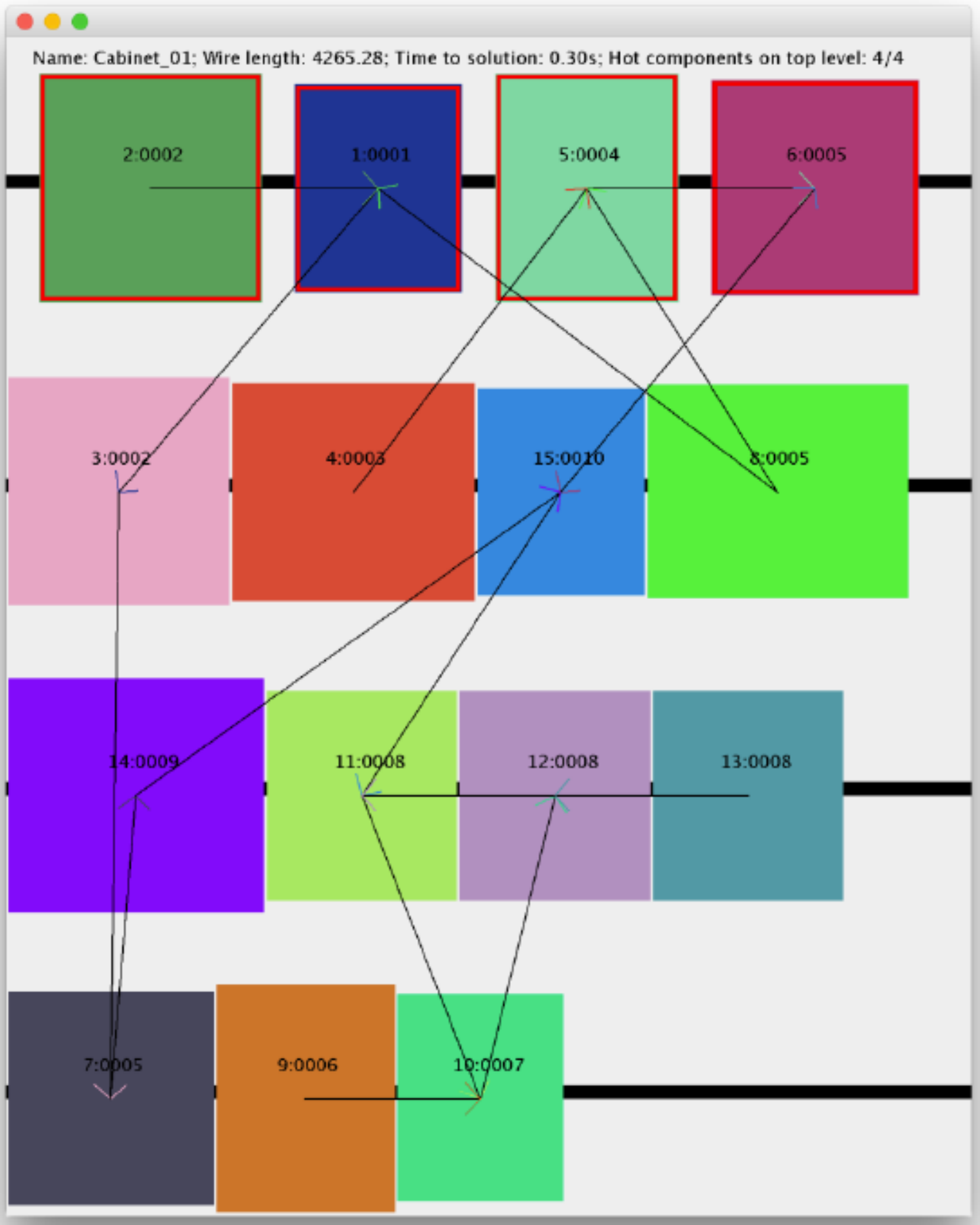}
		\caption{New component \#14 is wider and reconnected.}
		\label{fig:reconnected-component}
	\end{subfigure}	
	\caption{Interactive use of our implementation of cabinet layout optimization for layout reconfiguration. When a component is replaced, the optimization procedure quickly finds the new optimal cabinet layout.}
    \label{fig:reconfiguration}
\end{figure*}

Figure \ref{fig:initial-solution} depicts the optimized layout of the control cabinet described in Table \ref{tab:initial-description}. We may observe that the optimization procedure has placed all components with the property Is Hot equal to one (that is, components \#1, \#2, and \#5) on the top row of control cabinet. When component \#8 is replaced with a wider version, the layout re-optimization procedure took only 0.3 seconds and the result is visualised in Figure \ref{fig:wider-component}. Figure \ref{fig:hot-component} depicts the result of layout re-optimization for a new component \#6 that dissipates heat. Figure \ref{fig:reconnected-component} illustrates the scenario of replacing the component \#14  with a wider version and correcting the connection of component \#14 to other components.  

\section{Related work}
\label{sec:rw}

In this section, we discuss related work that addresses the issue of automatic layout generation. Additionally, examples of related work that use Simulated Annealing for optimization of various systems are highlighted. 

Merrell et al. \cite{Merrell:2010} propose an approach that is based on Bayesian networks for automatic generation of residential building layouts in the context of computer graphics applications (such as, computer games). Authors define a cost function that aims at avoiding layout anomalies, such as, ill-formed rooms or incompatibilities between floors. 

Pop et al. \cite{Pop:2008} surveyed various methods (such as, Simulated Annealing) for decentralized scheduling in Grid computing environments. Grid scheduling involves mapping of a collection of tasks to resources with the aim of minimizing the total execution time of all considered tasks. Compared to a game theory scheduling method, the average scheduling time per task of Simulated Annealing was lower.

Memeti et al. \cite{Memeti:2017,Memeti:2018a} use Simulated Annealing for optimization of DNA sequence analysis on heterogeneous computing systems that comprise a host with multi-core processors and one or more many-core devices. The optimization procedure aims at determining the number of threads, thread affinities, and DNA sequence fractions for host and device, such that the overall execution time of DNA sequence analysis is minimized.

Drexl and Nikulin \cite{Drexl:2008} use Pareto Simulated Annealing for solving the gate assignment problem in the context of an airport. Various aspects of airport operation are considered, such as, the total passenger walking distance, open flights, connection time, and gate assignment preferences. 

Yu et al. \cite{Yu:2017} propose to use Simulated Annealing for solving the hybrid vehicle routing problem. The aim is to minimize the total travel cost for hybrid vehicles that use fuel and electricity while considering the time limit, electric capacity, fuel capacity, locations of fuel stations and electricity charging stations. 

In contrast to the related work, we propose a method for multi-objective optimization of control cabinet layouts that is based on Pareto Simulated Annealing. Objectives of the optimization procedure are heat convection and wire length minimization. 

\section{Summary}
\label{sec:summary}

We have described an approach for customization of the multi-objective Simulated Annealing method for combinatorial optimization of control cabinet layout, which considers the heat convection and the total wire length used for interconnection of components. We used simulation to study the warm air flow within the control cabinet and determine the optimal position of components that generate heat during the operation. We observed that the best position for hot components is the top of control cabinet. We evaluated experimentally our approach using control cabinets with 14, 21, 41 components. For a cabinet with 14 components, trying about 4.7\% of all feasible configurations, required about 6.4 seconds and results with average 2.12x improvement of the wire-length in addition to the average 1.6x improvement of the heat-level. While an experienced engineer typically needs hours to plan the layout of a moderate size control cabinet, our solution generated a near-optimal layout within few seconds. We studied the interactive use of our implementation of control cabinet layout optimization for adapting the layout when a component is replaced. Re-optimization of the control cabinet under study using our solution took only about 0.3 seconds. 

Observed benefits of our solution:
\begin{itemize}
    \item generates much faster than the human cabinet layouts, 
    \item reduces time-to-solution from hours to seconds,
    \item minimizes total wire length,
    \item considers the heat convection by placing the hot components on the top.
\end{itemize}

Future work will compare the performance of Pareto Simulated Annealing with other meta-heuristics for combinatorial optimization of control cabinet layout. 

\section*{Acknowledgments} 
We are grateful to Robert Bick\"{o} and his colleagues at Yaskawa Nordic AB for providing an industrial perspective of the problem definition. We also thank Weibin Yu for discussions related to the implementation of Pareto Simulated Annealing. 



\end{document}